\documentclass[11pt]{SciPost}

\hypersetup{
    colorlinks,
    linkcolor={red!50!black},
    citecolor={blue!50!black},
    urlcolor={blue!80!black}
}

\usepackage{multirow}

\usepackage[bitstream-charter]{mathdesign}
\DeclareSymbolFont{usualmathcal}{OMS}{cmsy}{m}{n}
\DeclareSymbolFontAlphabet{\mathcal}{usualmathcal}

\fancypagestyle{SPstyle}{
\fancyhf{}
\lhead{\colorbox{scipostblue}{\bf \color{white} ~SciPost Physics}}
\rhead{{\bf \color{scipostdeepblue} ~Submission }}

\fancyfoot[C]{\textbf{\thepage}}
}

\usepackage{slashed}
\usepackage{graphicx}

\usepackage{soul}

\DeclareRobustCommand{\Sec}[1]{Sec.~\ref{#1}}

\DeclareRobustCommand{\Tab}[1]{Table~\ref{#1}}

\DeclareRobustCommand{\Fig}[1]{Fig.~\ref{#1}}

\DeclareRobustCommand{\Eq}[1]{Eq.~(\ref{#1})}

\DeclareRobustCommand{\InRef}[1]{Ref.~\cite{#1}}
\DeclareRobustCommand{\Refs}[1]{Refs.~\cite{#1}}

\newcommand{\be}{\begin{eqnarray}}
\newcommand{\ee}{\end{eqnarray}}

\allowdisplaybreaks

\usepackage{color}
\definecolor{darkblue}{rgb}{0,0,0.5}
\definecolor{darkgreen}{rgb}{0,0.5,0}

\begin{document}

\pagestyle{SPstyle}

\begin{center}{\Large \textbf{\color{scipostdeepblue}{
Systematic Interpretability and the Likelihood for Boosted Top Quark Identification
}}}\end{center}

\begin{center}\textbf{
Andrew J. Larkoski
}\end{center}

\begin{center}
\href{mailto:larkoa@gmail.com}{\small larkoa@gmail.com}
\end{center}

\section*{\color{scipostdeepblue}{Abstract}}
\textbf{\boldmath{%
Identification of boosted, hadronically-decaying top quarks is a problem of central importance for physics goals of the Large Hadron Collider.  We present a theoretical analysis of top quark tagging, establishing zeroth-order, minimal assumptions that should be satisfied by any purported top-tagged jet, like existence of three hard subjets, a bottom-tagged subjet, total mass consistent with the top quark, and a pairwise subjet mass consistent with the $W$ boson.  From these minimal assumptions, we construct the optimal discrimination observable, the likelihood ratio, for the binary discrimination problem of top quark-initiated versus bottom quark-initiated jets through next-to-leading order in the strong coupling.  We compare and compute corresponding signal and background efficiencies both analytically and from simulated data, validating an understanding of the relevant physics identified and exploited by the likelihood.
In the process, we construct a method for systematic interpretability of the likelihood ratio for this problem, and explicitly establish a hard floor on possible discrimination power.  These results can correspondingly be applied to understanding and interpreting machine learning studies of this problem.
}}

\vspace{\baselineskip}



\vspace{10pt}
\noindent\rule{\textwidth}{1pt}
\tableofcontents
\noindent\rule{\textwidth}{1pt}
\vspace{10pt}

\section{Introduction}

Machine learning has upended particle physics, especially in subfields of which central problems are particle identification, discrimination, and physical process determination \cite{Larkoski:2017jix,Kogler:2018hem,Guest:2018yhq,Albertsson:2018maf,Radovic:2018dip,Carleo:2019ptp,Bourilkov:2019yoi,Schwartz:2021ftp,Karagiorgi:2021ngt,Boehnlein:2021eym,Shanahan:2022ifi,Plehn:2022ftl,Nachman:2022emq,DeZoort:2023vrm,Zhou:2023pti,Belis:2023mqs,Mondal:2024nsa,Feickert:2021ajf,Larkoski:2024uoc,Halverson:2024hax}.  Because of the general, user-friendly, and accurate simulation software that is widely available, constructing, testing, and validating a machine learning architecture for a problem in particle physics has an extremely low barrier to entry, and correspondingly more and more techniques exist whose efficiency and efficacy are far beyond what could have been imagined a decade ago.  However, the overwhelmingly vast majority of machine learning techniques are only implemented on simulated data, and are therefore only as good as the data itself, and further, lack human interpretability for what physics the output exploits.  Especially for classification tasks, signal and background definitions are typically just taken from a user's process request in the simulation software.  At best, such class definitions are only well-defined to leading order in perturbation theory, and so drawing robust, theoretically sound conclusions from a machine learning study is challenging, if not impossible.

On the other hand, many binary discrimination problems in particle physics, and especially in jet physics, can be formulated and studied theoretically, in a systematic way.  By the Neyman-Pearson lemma \cite{Neyman:1933wgr}, the optimal discrimination observable is the likelihood ratio, and for appropriate problems of interest, this can constructed order-by-order in the perturbation theory of quantum chromodynamics (QCD) \cite{Larkoski:2023xam}.  Such a construction produces analytic, closed-form functional expressions for the likelihood on multi-particle phase space and whose description is systematically-improvable to whatever accuracy one is strong enough to calculate.  For such problems, foundational issues with interpretation of what a machine is learning are completely side-stepped, because one has complete control on the physics that is input into the theoretical analysis of the likelihood.

In this paper, we continue the program of \InRef{Larkoski:2023xam} which calculated the likelihood ratio and its discrimination power for the problem of distinguishing $H\to b\bar b$ decays from $g\to b\bar b$ fragmentation in highly-boosted jets.  The results there hinged on a systematic theoretical construction of the likelihood ratio $\hat {\cal L}(\Pi)$ as a function on phase space $\Pi$, where,
\begin{align}
\hat {\cal L}(\Pi) = \frac{p_b(\Pi)}{p_s(\Pi)} \approx \frac{p_b(\Pi_\text{NLO}|\Pi_\text{LO})}{p_s(\Pi_\text{NLO}|\Pi_\text{LO})}\,  \frac{p_b(\Pi_\text{LO})}{p_s(\Pi_\text{LO})}\,.
\end{align}
Here, $s$ and $b$ denote signal and background distributions, respectively, ``LO'' means leading-order in the strong coupling and ``NLO'' means next-to-leading order.  The ratio of leading-order distributions is independent of the coupling, while the ratio of conditional distributions at next-to-leading order and beyond has a Taylor series in $\alpha_s$, starting at $\alpha_s^0$.  In \InRef{Larkoski:2023xam}, it was shown that there is very little discrimination power at leading order for the $H\to b\bar b$ versus $g\to b\bar b$ problem, but starting at next-to-leading order, at which you first become sensitive to the color-singlet versus color-octet nature of the problem, the discrimination power is effectively unbounded, and is formally perfect in the infinite boost limit.

Here, we follow a similar line of analysis for the problem of boosted hadronically-decaying top quark identification.  In historical or machine learning analyses of this problem, the background is often taken to simply be jets initiated by light QCD partons, see, e.g., \Refs{Butter:2017cot,Kasieczka:2019dbj,Chakraborty:2020yfc,Gong:2022lye,Qu:2022mxj,Baldi:2022okj,Bogatskiy:2022czk,Gambhir:2024dtf}, but as an experimentally-viable procedure, this includes many events that could really never be identified as a top quark anyway.  A top quark essentially always decays into a bottom quark and a $W$ boson, so a natural requirement for the background is that there is an identified bottom quark or hadron in the jet.  Along with this $b$-tagging, we identify other minimal requirements that are trivially satisfied on top quark jets, at least in the narrow-width approximation, and that dramatically reduce the background that needs to be considered.  Here, we focus on binary discrimination of hadronically-decaying top quark jets from jets initiated by bottom quarks, that further subsequently fragment into at least three partons.

With this groundwork established, from the exact expressions for leading-order matrix elements in the collinear limit, we construct the likelihood ratio and demonstrate that there is significant discrimination information encoded in the kinematic distributions of the three subjets present at leading order.  Sensitivity to emissions and especially the flow of color in the jets is first present at next-to-leading order, but top and bottom quarks are both color triplets, and so there is rather limited discrimination power at next-to-leading order and beyond.  This suggests that there is a fundamental lower bound or discrimination floor to background rejection for this problem, effectively determined by the efficacy of $b$-tagging and the information encoded in kinematics at leading order.

This theoretical analysis is then tested in simulation, where we validate that the analytical expression for the leading-order likelihood we derived from matrix elements remains a powerful discriminant in simulation.  General-purpose parton shower generators employ a number of assumptions or approximations, and these may be especially relevant for drawing quantitative conclusions.  Parton showers typically generate emissions through subsequent $1\to 2$ splittings, and the corresponding matrix element is distinct from the complete $1\to 3$ splitting function, and we show actually results in reduced discrimination power.  Further, color is typically only managed at leading-order in the large number of colors $N_c\to \infty$ limit, which reduces the possible dipole connections between particles and how radiation is subsequently emitted.  Techniques such as color reconnections may assuage some of the limitations of the leading color approximation, but also may randomize color in a way that isn't necessarily represented in the matrix element.  The limited improvement in discrimination power at next-to-leading order by accounting for color connections and soft gluon emission in simulation suggests further work is needed to validate the physics of the parton shower, or may require simply using a more accurate shower.

The outline of this paper is as follows.  In \Sec{sec:minass}, we establish the minimal assumptions that we employ throughout this paper for defining what a ``top quark jet'' could possibly be.  In this section, we also provide a parametric estimate of the background rejection that follows from these assumptions, and find qualitative agreement with rejection rates established in machine learning studies, i.e., from \InRef{Kasieczka:2019dbj}.  In \Sec{sec:loanal}, we present the detailed analysis of the likelihood ratio at leading order in the strong coupling, working in the highly-boosted, collinear limit.  We construct a simple discrimination observable that closely approximates the likelihood, and calculate its distribution on signal and background jets.  In \Sec{sec:nloanal}, we continue the theoretical analysis to next-to-leading order, but restrict our focus to just the kinematics of soft gluon emission.  This motivates construction of an infrared and collinear (IRC) safe color flow observable directly from the eikonal matrix elements.  These analytic results are then tested in simulation in \Sec{sec:toptagprac}, where the efficacy of the analytic observables we constructed is verified.  We conclude in \Sec{sec:concs}, summarizing our results and looking forward to more studies, tests and validation that can be done to fully understand and interpret top quark tagging as a human.

\section{Minimal Working Assumptions for Identifying a Top-Tagged Jet}\label{sec:minass}

For the rest of this paper, we will work with a minimal set of assumptions for a highly-boosted jet to possibly be tagged as originating from hadronic top quark decay.  The assumptions we use are:
\begin{enumerate}
\item The mass of the jet $m_J$ is around the mass of the top quark, $m_J \sim m_t$.
\item The jet has at least three hard subjets, where ``hard'' means that the corresponding splitting scale between pairs of subjets is comparable to the mass of the top quark itself.
\item Exactly one of the hard subjets contains a bottom hadron, or is ``$b$-tagged''.
\item A pair of the non $b$-tagged subjets have an invariant mass that is around the mass of the $W$ boson.
\end{enumerate}
For assumption 1, we correspondingly assume that there is minimal contamination radiation in the jet that could bias the jet mass measurement significantly larger than $m_t \sim 172$ GeV.  In the analytical study, we will work in the narrow-width approximation and force the jet mass to be precisely the mass of the top quark, $m_J = m_t$.  In the simulation studies of \Sec{sec:toptagprac}, we must relax this to a window about the top mass, $m_J \in[m_t-m_-,m_t+m_+]$, where $m_-,m_+$ are relatively small mass scales compared to the mass of the top.

For assumption 2, we consider reclustering the jet with a sequential jet algorithm, terminating with three subjets.  Demanding that these subjets are hard can be enforced by requiring that the pairwise invariant mass between any two of these subjets $i,j$ is larger than some cut, $m_{ij} > m_\text{cut}>0$.  Assumptions 3 and 4 are then further constraints on these subjets.  Practically implementing $b$-tagging in our calculations and simulations will be as simple as perfectly identifying the flavor of the appropriate subjet and tagging it as bottom flavor if there exists a bottom quark (or hadron, in simulation) in it.  Experimentally, $b$-tagging is extremely accurate and efficient \cite{ATLAS:2022qxm,CMS:2023tlv}, so this is well-justified.  On the other hand, defining the flavor of a (sub)jet is rather subtle, and the simple approach we use here cannot be generalized to arbitrarily high orders in perturbation theory.  Significant recent work has been devoted to developing theoretically well-defined jet flavor definitions, e.g., \Refs{Banfi:2006hf,Caletti:2022hnc,Caletti:2022glq,Czakon:2022wam,Gauld:2022lem,Caola:2023wpj}, but we won't need to worry about the subtleties to the order that we work.  Finally, for assumption 4, in our analytical study we work again in the narrow-width approximation for the $W$ boson and require that the invariant mass of the two non-$b$-tagged subjets $1,2$ is exactly the $W$ mass, $m_{12} = m_W$.  In simulation, their mass is within a window about the $W$ mass, $m_{12}\in[m_W-m_{W-},m_W+m_{W+}]$, where $m_{W-},m_{W+}$ are relatively small mass scales compared to the mass of the $W$.

With this set of assumptions, there are two dominant final state particle configurations possible for the QCD background at leading-order in perturbation theory.  Requiring at least three hard subjets forces there to be at least three particles, and one of those particles must be a bottom quark.  In the highly-boosted limit, in which the energy of the jet $E$ is much larger than its mass, $E/m_t \gg 1$, the dominant background is described by the collinear fragmentation of a high-energy parton, which must itself be a bottom quark.  Therefore, the possible leading-order background particle configurations are $bq\bar q$, for some non-bottom quark $q$, and $bgg$.  An analytical treatment of the $bgg$ background is rather subtle because, even after imposing the top and $W$ mass constraints, there are still residual collinear and soft divergences that must be regulated through clustering the jet with a finite jet radius $R < \infty$ and further enforcing the minimal pairwise mass cut, $m_{ij}>m_\text{cut}$.  As such, an analysis that includes the $bgg$ background is sensitive to at least these two parameters, which complicates analytical calculations and conclusions drawn from them.

By contrast, the $bq\bar q$ background has neither collinear nor soft divergences once the top and $W$ mass constraints are imposed and so to produce finite predictions, no additional constraints need to be imposed.  Correspondingly, this means that discrimination of hadronic top decay $t\to bq\bar q'$ from the collinear fragmentation $b\to bq\bar q$ in the sufficiently high-energy limit is independent of the precise parameters used to cluster and find jets.  Because of this simplicity, we will restrict our theoretical analysis to this $b\to bq\bar q$ background final state, but will include all QCD final states consistent with the assumptions above in our simulations later.  While we will not study it more here, this set-up makes it clear that a quark versus gluon subjet discriminant, e.g., historical studies of \Refs{Gallicchio:2011xq,Gallicchio:2012ez,Larkoski:2013eya,Gras:2017jty}, may have significant impact in improving discrimination of top quark decays from QCD processes.

The set of minimal assumptions we use in this paper are primarily motivated by the theoretical analysis, to ensure that the corresponding optimal discrimination observable, the likelihood ratio, is infrared and collinear safe and so can be calculated in perturbation theory.  However, similar requirements have long been employed in top tagging studies, e.g., \Refs{Kaplan:2008ie,Thaler:2008ju,Ellis:2009su,Ellis:2009me,Almeida:2010pa,Thaler:2010tr,Jankowiak:2011qa}, and even in recent experimental results from ATLAS and CMS related requirements are imposed or effectively identified through a machine learning analysis, e.g., \Refs{ATLAS:2018wis,ATLAS:2023jdw,CMS:2020tvq,CMS:2020poo}.  Nevertheless, it is interesting to consider if these assumptions could be further relaxed, especially if there exists some implicit, but as-of-yet unidentified, bias that exists within these assumptions.  We leave this question to future work.

\subsection{Estimate of Nominal Signal and Background Efficiencies}\label{sec:nomratesest}

From these initial cuts that define a candidate top quark jet, we can estimate the probability that a given background jet will pass these criteria and need to be further analyzed.  First and foremost, the requirement of a $b$-tagged subjet effectively requires that a bottom quark is produced in the hard scattering at leading order.  This isn't quite true because bottom quarks can be produced in the parton shower itself, but production of bottom quarks in a parton shower always occurs in relatively collinear $b\bar b$ pairs and is suppressed by at least a power of the coupling, $\alpha_s$.  So, to leading approximation, we will assume that the background jets that could possibly be candidate top quarks are initiated by bottom quarks at short distances.

At the sufficiently high energies in which we work in this paper, both the mass of the bottom quark and the mass of the top quark are very small compared to the relevant jet energy scale, so to leading approximation in the high-boost limit, we can assume they are massless for estimating inclusive production.  Further, both the bottom and top quark masses are large compared to the proton's mass, and so neither exist as potential colliding partons at a hadron collider.  Therefore, at least in QCD processes at sufficiently high energies, bottom and top quark production is otherwise identical, and so the ratio of their inclusive cross sections will approach unity.  Concretely, if we consider pair production at the LHC, $pp\to b\bar b$ and $pp\to t\bar t$, the limit of the ratio of inclusive cross sections is 
\begin{align}
\lim_{p_\perp \to\infty}\frac{\sigma_{pp\to t\bar t}}{\sigma_{pp\to b\bar b}} = 1\,,
\end{align}
where $p_\perp$ is the characteristic jet transverse momentum.

Now, given that a jet contains a bottom quark (or bottom hadron experimentally), we would like to estimate the probability that such a jet further passes the subjet mass constraints.  In the high-boost limit, any sufficiently large jet radius is irrelevant because all decay or fragmentation products will be contained within the jet, and so a signal jet from hadronic top quark decay will always pass the total jet and subjet mass cuts.  By contrast, a background bottom quark-initiated jet is highly constrained with the additional cuts and is rather unlikely to pass the cuts.  The probability that a jet passes the cuts given that it was initiated by a $b$ quark can be estimated from integrating the collinear splitting function squared matrix element $|{\cal M}|^2$ over the appropriate phase space $d\Pi$:
\begin{align}
p(\text{passes cuts}|b \text{ jet}) \sim d\Pi\, |{\cal M}|^2\,.
\end{align}
We will present estimates of the matrix element and the phase space in turn.

First for the matrix element, the final state must contain at least three particles so that three hard subjets are resolved.  There are then three contributions to bottom quark fragmentation to three partons: $b\to bq\bar q$ where $q$ is a non-bottom quark, $b\to bgg$ where gluons are emitted like photons (the Abelian contribution), and $b\to bgg$ where the gluons are correlated (the non-Abelian contribution).  The $b\to bq\bar q$ contribution is proportional to the number of non-bottom quarks, while the Abelian and non-Abelian contributions are proportional to the fundamental and adjoint quadratic Casimirs $C_F$ and $C_A$ of QCD color, respectively.  The production of three final state particles can be modeled as subsequent $1\to 2$ processes, and the first such splitting has a color factor $C_F$ because a gluon must be emitted from a bottom quark.  Demanding that the total jet mass is $m_t$, the initial propagator of the splitting scales like $1/m_t^2$, and then demanding that the $W$ boson mass $m_W$ is produced in the second $1\to 2$ splitting sets the scale of the second propagator to be $1/m_W^2$.  The production of two additional particles requires two factors of the strong coupling $\alpha_s$, and so combining these factors, the matrix element is approximately
\begin{align}
|{\cal M}|^2 \sim\left(\frac{\alpha_s}{2\pi}\right)^2\frac{C_F}{m_t^2 m_W^2}\left(
(n_f-1)T_R + C_F+C_A
\right)\,.
\end{align}
In QCD, $T_R = 1/2$, $C_F  =4/3$, and $C_A = 3$.  $n_f-1$ is the number of non-bottom active quarks, which we take to be 4.

The volume of phase space $d\Pi$ can be determined sequentially through two $1\to 2$ splittings.  Two-body collinear phase space is
\begin{align}
d\Pi_2 \sim dz\, ds\,,
\end{align}
where $z$ is an energy fraction and $s$ is the total invariant mass of the pair of particles.  For the initial splitting, which would emulate the $t\to bW$ decay in the top quark, the bottom is effectively massless, while the $W$ boson is massive.  If one particle has mass $m_W^2$ and the other is massless, then the range of the energy fraction of the massless particle is $z\in[0,1-m_W^2/s]$.  Therefore, the volume of $b\to bW$ phase space is 
\begin{align}
d\Pi_2^{(1)} \sim \frac{m_t^2-m_W^2}{m_t^2}\, dm_t^2\,
\end{align}
where $s\sim m_t^2$ and $dm_t^2$ is the size of the window about the top mass.  Next, we need to consider the $W$ decay to massless partons.  Their energy fraction is now allowed to range over all $z\in[0,1]$, and the window about the $W$ mass is $dm_W^2$.  Then, the total volume of this secondary phase space is approximately
\begin{align}
d\Pi_2^{(2)} \sim dm_W^2\,.
\end{align}
The volume of the total phase space is then the product of these two estimates, 
\begin{align}
d\Pi\sim d\Pi_2^{(1)}\,d\Pi_2^{(2)} \sim \frac{m_t^2-m_W^2}{m_t^2}\, dm_t^2\, dm_W^2\,.
\end{align}

Now, putting this all together and multiplying the matrix element estimate by the phase space volume estimate, the probability that the fragmentation of an initial bottom quark passes the subjet and mass cuts is
\begin{align}\label{eq:bsplitest}
p(\text{passes cuts}|b\text{ jet}) \sim \frac{m_t^2-m_W^2}{m_t^2}\, \frac{dm_t^2}{m_t^2}\, \frac{dm_W^2}{m_W^2}\,\left(\frac{\alpha_s}{2\pi}\right)^2C_F\left(
(n_f-1)T_R + C_F+C_A
\right)\,.
\end{align}
To get a sense for the scale of this probability, note that in practice, the mass-dependent prefactors are roughly order-1.  Mass windows about the top or $W$ mass are typically a relatively large fraction of the mass, like 30\% or so, and the mass of the $W$ boson is about 45\% of the mass of the top.  So, the order-of-magnitude scaling of this probability is controlled by the coupling and color factors.  Because of the explicit mass scales imposed on the jet, the coupling $\alpha_s$ will be evaluated at the top or $W$ mass, at which $\alpha_s \sim 0.1$.  Accounting for the factors of $2\pi$, the coupling suppression itself is on the order of $10^{-4}$.  The product of color factors is about 10, and so a rough estimate of the background efficiency from these nominal cuts is
\begin{align}
p(\text{passes cuts}|b\text{ jet}) \sim 10^{-3}\,.
\end{align}

Note that given an initial $b$ jet, the constraints we identified were exclusively kinematic, the number of hard subjets and corresponding invariant masses, and so this logic can be applied to top tagging at this level on any initial jet.  Color factors may slightly change depending on if the jet is initiated by a quark or gluon, but these will be relatively small effects to a general estimate.  Thus, we expect that the probability that any jet initiated by a light QCD parton looks like that of a hadronic top quark decay is roughly $10^{-3}$, or, that the rejection rate of a QCD jet for top tagging is approximately $10^3$.  This characteristic $10^{3}$ kinematics rejection rate is also what has been observed in detailed machine learning studies of top taggers from \InRef{Kasieczka:2019dbj} in which only kinematic information was exploited.  In that study, because no flavor information was used, background jets were selected from simulated inclusive $pp\to $ dijet production at the LHC.  The ratio of cross sections of inclusive dijet production to $t\bar t$ production at high energies is very large,
\begin{align}
\frac{\sigma_{pp\to jj}}{\sigma_{pp\to t\bar t}} \sim 500\,,
\end{align}
which we have estimated for jets with transverse momentum around 1 TeV at the 13 TeV LHC with MadGraph v3.6.0 \cite{Alwall:2014hca}.  Thus, applying the nominal kinematic cuts on QCD jets initiated by light partons renders the number of background and signal jets in a sample comparable, and then one can study the jets in more detail to further reduce background.

\section{Leading-Order Analysis}\label{sec:loanal}

In this section, we present the analysis of discrimination of boosted, hadronic top quark decay from massive jets initiated by light QCD partons.  As discussed in the previous section, we will concretely only consider binary discrimination of the processes $t\to b q\bar q'$ from $b\to bq\bar q$ in the limit in which the energy of the jets is much larger than their mass, $m_t/E \ll 1$.  As such, the signal and background processes are highly collinear, and so are to good approximation described by collinear fragmentation to leading power in $m_t/E$.  Differential three-body collinear phase space for this problem can be expressed as \cite{Gehrmann-DeRidder:1997fom,Ritzmann:2014mka}
\begin{align}
d\Pi_3 &=\frac{4}{(4\pi)^5} \frac{ds_{q\bar q}\, ds_{bq}\, ds_{b\bar q}\,dz_q\, dz_{\bar q}\, dz_b\, \delta(1-z_q-z_{\bar q}-z_b)}{\sqrt{4z_qz_{\bar q} s_{bq}s_{b{\bar q}}-(z_b m_W^2-z_q s_{b\bar q}-z_{\bar q}s_{bq})^2}}\,\delta(m_t^2-m_W^2-s_{bq}-s_{b\bar q})\,\delta(s_{q\bar q}-m_W^2)\,.
\end{align}
Here, $s_{ij}$ is the invariant mass of partons $i,j$ and $z_i$ is the energy fraction of parton $i$.  The rightmost two $\delta$-functions impose the total top mass constraint and the $W$ mass subjet constraint, respectively.  There is also an implicit positivity constraint on the discriminant that appears in the square-root.

The leading-order distribution for top decay $t\to bq\bar q'$ on this phase space (as calculated in the narrow-width approximation from the left-handed weak decay of the top) is
\begin{align}\label{eq:topweak}
p_s(\Pi_3) = {\cal N}_s\,
\frac{s_{bq}(m_W^2+s_{b\bar q'})}{m_t^4}={\cal N}_s\,
\frac{s_{bq}(m_t^2-s_{bq})}{m_t^4}\,,
\end{align}
where ${\cal N}_s$ is a normalization factor such that it integrates to 1 on phase space $\Pi_3$.  The background process, $b\to bq\bar q$, is described by the $1\to 3$ collinear splitting function, where \cite{Campbell:1997hg,Catani:1999ss}
\begin{align}\label{eq:qcdfo}
p_b(\Pi_3) &= {\cal N}_bC_F(n_f-1)T_R\left(
\frac{\alpha_s}{2\pi}
\right)^2\frac{(4\pi)^4}{2m_t^2m_W^2}\\
&\hspace{1cm}\times\left[
-\frac{\left( z_q(m_W^2+2s_{b\bar q})-z_{\bar q}(m_W^2+2s_{bq}) \right)^2}{(1-z_b)^2 m_t^2m_W^2}+\frac{4z_b+(z_q-z_{\bar q})^2}{1-z_b}+1-z_b-\frac{m_W^2}{m_t^2}
\right]\,,\nonumber
\end{align}
where ${\cal N}_b$ is a normalization factor.

With these explicit expressions for the differential phase space and splitting function, we can more precisely calculate the rate for the $b\to bq\bar q$ fragmentation to pass the pre-established cuts.  Setting the normalization factor ${\cal N}_b = 1$ so that the resulting prediction is a rate relative to inclusive bottom quark production, we can numerically integrate over phase space to find
\begin{align}
\int d\Pi_3\, p_b(\Pi_3) \approx\frac{m_t^2-m_W^2}{m_t^2}\, \frac{dm_t^2}{m_t^2}\, \frac{dm_W^2}{m_W^2}\,\left(\frac{\alpha_s}{2\pi}\right)^2C_F
(n_f-1)T_R \,\frac{14.5}{4\pi}\,.
\end{align}
Here, we have set $m_t = 172$ GeV and $m_W = 80$ GeV, and the error on the result of the numerical integral, $14.5$, is on the order of the last quoted digit.  Note that $14.5/4\pi \approx 1.15$, and so the estimate we presented in \Eq{eq:bsplitest} for this color channel is, perhaps, surprisingly accurate.

Note that the background distribution is symmetric in quark and anti-quark, because the gluon is a vector boson, while the top decay distribution is not symmetric, because the weak force is left-handed.  Experimentally, if we only have access to energy deposits in the calorimetry, a quark and anti-quark are indistinguishable, so we should symmetrize the top quark decay distribution, where
\begin{align}
p_{s,\text{sym}}(\Pi_3) = \frac{{\cal N}_s}{2}\left[
\frac{s_{bq}(m_W^2+s_{b\bar q'})}{m_t^4}+\frac{s_{b\bar q'}(m_W^2+s_{bq})}{m_t^4}\right]=\frac{{\cal N}_s}{2}\,
\frac{m_W^2(m_t^2-m_W^2)+2s_{bq}s_{b\bar q'}}{m_t^4}\,.
\end{align}
For concrete comparison, we will study both the original weak decay distribution, \Eq{eq:topweak}, and this symmetrized distribution to establish the information that is lost for discrimination when symmetrized.  With access to more information, like the charged particle content, it is possible to discriminate the quark and anti-quark within top quark decay at an aggregate, statistical level \cite{Dong:2024xsg}.  One of the most useful single observables for this is the jet charge \cite{Field:1977fa,Krohn:2012fg,Waalewijn:2012sv,Fraser:2018ieu,Kang:2023ptt}, but we leave explicit inclusion of jet charge in a theoretical analysis to future work.  Here, we will just use the left-handed and symmetrized decay of the top as the extreme bounds of ignorance or knowledge of the flavor of the additional quarks in the jet.

Within a typical workhorse parton shower event generator, like Pythia, Herwig, or Sherpa \cite{Bierlich:2022pfr,Bahr:2008pv,Bellm:2015jjp,Sherpa:2019gpd}, particles are generated in the parton shower through strongly-ordered sequential $1\to 2$ splittings.  Therefore, without otherwise matching to fixed-order, the three particles in the background jet in simulation would not have a distribution on phase space of \Eq{eq:qcdfo}, but rather only the leading contribution when assuming that the invariant mass of the $q\bar q$ pair is parametrically smaller than that of the total jet mass, $m_W \ll m_t$.  Of course, this isn't necessarily a good approximation, as $m_W/m_t \sim 0.47$, but is another interesting limit to consider for bounding discrimination power.  In this strongly-ordered limit, the collinear $1\to 3$ splitting function becomes
\begin{align}
p_{b,\text{SO}}(\Pi_3) = {\cal N}_{b,\text{SO}}C_F(n_f-1)T_R\left(
\frac{\alpha_s}{2\pi}
\right)^2\frac{(4\pi)^4}{2m_t^2m_W^2}\left[
\frac{1+z_b^2}{1-z_b}\,\frac{z_q^2+z_{\bar q}^2}{(z_q+z_{\bar q})^2} - \frac{8z_bz_qz_{\bar q}}{(1-z_b)^3}\cos(2\phi)
\right]\,.
\end{align}
The first term in the brackets is the product of $1\to 2$ splitting functions $b\to bg$ and $g\to q\bar q$, respectively, and the second term, with the $\cos(2\phi)$ factor, is the interference term between the two helicity states of the collinear gluon.  $\phi$ is the azimuthal angle of the $q\bar q$ pair about the $b$ quark, which satisfies the law of cosines
\begin{align}
z_q s_{b\bar q} = z_bm_W^2 + z_{\bar q}s_{bq}-2m_W\sqrt{z_b z_{\bar q} s_{bq}}\cos\phi\,.
\end{align}
Leading-logarithmic parton showers do not necessarily correctly describe this interference term, but we will include it for self-consistency of our results.  Nevertheless, the correct inclusion or incorrect exclusion of this term may affect the quantitative results that we present later, but that could only be addressed with a full next-to-leading logarithmic shower, e.g., \Refs{Hamilton:2020rcu,Forshaw:2020wrq,Nagy:2020rmk,Herren:2022jej,Preuss:2024vyu}.

\subsection{Discrimination with the Likelihood Ratio}\label{sec:lolike}

Given these signal and background distributions on phase space, we can then construct the optimal discrimination observable, the likelihood ratio, by the Neyman-Pearson lemma.  We will consider three likelihood ratios.  First, the likelihood ratio between the full $1\to 3$ $b\to bq\bar q$ splitting function and the left-handed top quark decay, where
\begin{align}
{\cal L} \equiv \frac{p_b(\Pi_3)}{p_s(\Pi_3)}\,.
\end{align}
As mentioned above, to implement this observable in a realistic analysis requires quark and anti-quark flavor identification which is likely not possible, but remains a useful bound for comparison.  The second likelihood ratio we consider is that with the top quark decay symmetrized over the final state $q$ and $\bar q'$, where
\begin{align}
{\cal L}_\text{sym} \equiv \frac{p_b(\Pi_3)}{p_{s,\text{sym}}(\Pi_3)}\,.
\end{align}
The third likelihood ratio we consider consists of the strongly-order bottom quark splitting function and the symmetrized top quark decay, where
\begin{align}
{\cal L}_\text{SO} \equiv \frac{p_{b,\text{SO}}(\Pi_3)}{p_{s,\text{sym}}(\Pi_3)}\,.
\end{align}
To calculate the distribution of these likelihood ratios on signal and background, for consistency we use the corresponding signal and background distributions represented in the likelihood.

Especially with the full splitting function, these likelihood ratios are complicated functions of the phase space variables, and we would like to construct a compact, yet powerful, discrimination observable that closely approximates the likelihoods for implementation in analysis.  Motivated by the strongly-ordered splitting function and the symmetrized top quark decay, we consider the observable ${\cal O}_\text{LO}$, where
\begin{align}\label{eq:loobs}
{\cal O}_\text{LO} \equiv \frac{m_W^2}{m_t^2}\frac{1}{1-z_b}\frac{m_t^4-m_W^4}{2\left(m_W^2(m_t^2-m_W^2)+2s_{bq}s_{b\bar q}\right)}\,.
\end{align}
Factors of the top and $W$ mass are there simply to limit the range of this observable.  The right-most factor is the inverse of the symmetrized top decay distribution, while the $1/(1-z_b)$ factor is a component of the background distribution.  This factor is large when the bottom quark takes most of the energy of the jet, which is likely because the emission of a gluon in the background distribution has a soft divergence.  Further, as constructed, this observable is rather ignorant to the secondary splitting at the $W$ mass scale, and so should generalize to the fragmentation process $b\to bgg$ more readily than the specific likelihood ratio formed from the $b\to bq\bar q$ splitting function.

\begin{figure}[t]
\begin{center}
\includegraphics[width=0.45\textwidth]{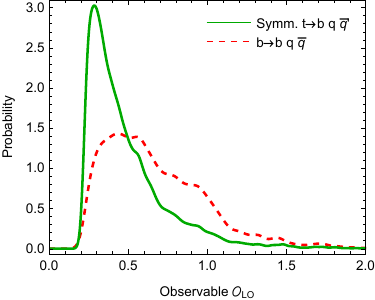}\ \ \ 
\includegraphics[width=0.45\textwidth]{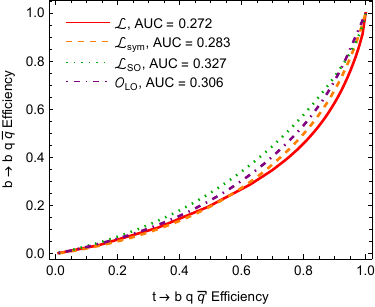}
\caption{\label{fig:loplots}
Left: Distribution of the observable ${\cal O}_\text{LO}$ on symmetrized top quark decay (solid green) and $b\to bq\bar q$ collinear fragmentation (dashed red) at leading order.  Right: ROC curves of the three likelihood ratios constructed from the left-handed top quark decay, the symmetrized top quark decay, the full $b\to bq\bar q$ splitting function, and the strongly-order $b\to bq\bar q$ splitting function.  Also plotted is the ROC curve from implementing a cut on the observable ${\cal O}_\text{LO}$.  For each of the observables, we also list the area under the ROC curve (AUC).
}
\end{center}
\end{figure}

At left in \Fig{fig:loplots}, we plot the distribution of this observable ${\cal O}_\text{LO}$ as calculated on the symmetrized top quark decay distribution, $p_{s,\text{sym}}(\Pi_3)$, and the full $1\to 3$ splitting function, $p_b(\Pi_3)$.  We evaluate these distributions by numerical Monte Carlo integration over the corresponding phase space distributions.  The signal distribution is peaked at small values, as expected because when the signal distribution is large, the observable ${\cal O}_\text{LO}$ is small.  By contrast, the background distribution is spread over a larger range.  At right in \Fig{fig:loplots}, we show the resulting signal versus background efficiency plots, or receiver operating characteristic (ROC) curves, for the three likelihood ratios and the observable ${\cal O}_\text{LO}$.  Better discrimination corresponds to the lower-right region of the plot.  Not surprisingly, the most effective discriminant is the likelihood that contains the left-handed top quark decay, but the symmetric decay is not much less discriminant.  Perhaps the most interesting feature of this plot is that the likelihood ratio of the strongly-ordered background distribution is the worst discriminant among what we study here.  This could illustrate a fundamental limitation of modeling discrimination with a strongly-ordered $1\to 2$ parton shower simulation program, especially for processes that involve three or more particles at leading order. 

\section{Next-to-Leading Order Analysis}\label{sec:nloanal}

We now move to analysis of the likelihood ratio for top quark discrimination at next-to-leading order.  As derived in \InRef{Larkoski:2023xam}, the likelihood ratio can be expanded through next-to-leading order in perturbation theory and takes the form:
\begin{align}
{\cal L} = \frac{p_b(\Pi_\text{LO})}{p_s(\Pi_\text{LO})}\left[
1+\frac{\alpha_s}{2\pi}\left(
\int d\Pi_\text{NLO}\left(
p_s(\Pi_\text{NLO})-p_b(\Pi_\text{NLO})
\right)+\frac{p_b(\Pi_\text{NLO})}{p_b(\Pi_\text{LO})}-\frac{p_s(\Pi_\text{NLO})}{p_s(\Pi_\text{LO})}
\right)+\cdots
\right]\,.
\end{align}
Note that the overall factor on the left is the leading-order likelihood ratio,
\begin{align}
{\cal L}_\text{LO} = \frac{p_b(\Pi_\text{LO})}{p_s(\Pi_\text{LO})}\,,
\end{align}
and the ${\cal O}(\alpha_s)$ term in the square brackets is the next-to-leading order correction to the likelihood.  $\Pi_\text{LO}$ is leading-order phase space, and the next-to-leading order contribution consists of two components: a normalization term (the integral over next-to-leading order phase space $\Pi_\text{NLO}$), and the difference of background and signal distributions, that vary over next-to-leading order phase space.  This likelihood ratio, its distribution on signal and background, and the corresponding ROC curve can be calculated with the next-to-leading order distributions for top decay and $1\to 4$ collinear splitting functions \cite{DelDuca:2019ggv,DelDuca:2020vst}.  The general procedure was outlined in \InRef{Larkoski:2023xam}, but even for the simpler case of $H\to b\bar b$ decays versus $g\to b\bar b$ splitting, the next-to-leading order analysis was extensive.  Our approach here will then not be to derive complete, exhaustive results to this order, but to identify relevant approximations that inform the construction of better discrimination observables that are sensitive to physics at next-to-leading order and beyond.

To this goal, we will just focus on the kinematic dependence at next-to-leading order, and ignore the normalization factor.  For top quark decay versus $b\to bq\bar q$ fragmentation, this difference term can be expressed as \cite{Larkoski:2024uoc}
\begin{align}\label{eq:likenlocolor}
\frac{p_b(\Pi_\text{NLO})}{p_b(\Pi_\text{LO})}-\frac{p_s(\Pi_\text{NLO})}{p_s(\Pi_\text{LO})}= -(4\pi)^2\sum_{\text{LO partons }i,j}\left({\mathbf T}^{(b)}_i\cdot{\mathbf T}^{(b)}_j-{\mathbf T}^{(s)}_i\cdot{\mathbf T}^{(s)}_j\right)\,\frac{s_{ij}}{s_{ik}s_{kj}}+\text{non-singular}\,.
\end{align}
Here, we have explicitly written out the most singular contribution, corresponding to soft gluon emission off of the hard partons at leading order.  For these signal and background processes, collinear contributions exactly cancel because the particles at leading order are identical, and so undergo the same collinear splittings.  Terms that are non-singular in the soft emitted gluon limit are suppressed and will not be considered here.  In this expression, ${\mathbf T}^{(b)}_i$ (${\mathbf T}^{(s)}_i$) is the color matrix of background (signal) event particle $i$ and $s_{ij}$ is the invariant mass of particles $i$ and $j$.  $k$ is the soft gluon emitted at next-to-leading order.  To identify kinematic dependence of the likelihood at next-to-leading order, we then need to evaluate the color matrix products.

As always in this paper, we are working in the highly boosted limit, in which the jet is collimated, or, equivalently, angles between pairs of particles within the jet are parametrically smaller than angles to particles outside the jet, in the rest of the event.  Further, the jets we are considering are initiated by colored partons, and so have color correlations with the rest of the event, by global color conservation.  So, in addition to the particles in the jet that can emit a soft gluon, we also need to include the particle representing the rest of the event off of which the jet recoils, which we call $\bar n$.  In the collinear limit of the jet, $\bar n$ is in the opposite direction of the jet and carries opposite net color to that of the jet.

With this set-up, we can then calculate the soft gluon emission contribution to the top quark decay.  In the narrow-width approximation, there is no color mixing between the color-singlet $W$ boson and the bottom quark.  Only the bottom quark couples to the rest of the event, and so the only non-zero color factors are
\begin{align}
{\bf T}_{\bar n}\cdot {\bf T}_b ={\bf T}_q\cdot {\bf T}_{\bar q'} = -C_F\,,
\end{align}
with all others 0.  The contribution to the likelihood from signal is then
\begin{align}
\frac{p_s(\Pi_\text{NLO})}{p_s(\Pi_\text{LO})} \supset -(4\pi)^2\sum_{i,j}{\bf T}^{(s)}_i\cdot {\bf T}^{(s)}_j\,\frac{s_{ij}}{s_{ik}s_{jk}} &= 2(4\pi)^2 C_F\left(
\frac{m_W^2}{s_{qk}s_{\bar q k}} + \frac{z_b}{z_k s_{bk}}
\right)\,.
\end{align}
One particularly interesting thing about this is that it is actually symmetric in the quark and anti-quark from $W$ decay, even though the leading-order matrix element is not.  However, this contribution is conditioned on the kinematics at leading-order, and so the emitted gluon only knows about the color connections, and nothing about the likelihood that the leading order particles arrange themselves in a particular configuration.

The soft gluon emission contribution to background is significantly more complicated, but can be evaluated by completeness relations of the SU($N$) generators (see, e.g., \InRef{Dixon:1996wi} for details).  The resulting expression for soft gluon emission is \cite{DelDuca:2019ggv}
\begin{align}
\frac{p_b(\Pi_\text{NLO})}{p_b(\Pi_\text{LO})}&\supset 2(4\pi)^2\left[
C_F \left(
\frac{m_W^2}{s_{qk}s_{\bar qk}}+\frac{2s_{b\bar q}}{s_{bk}s_{\bar qk}}-\frac{2s_{bq}}{s_{bk}s_{qk}}-\frac{2z_{\bar q}}{z_k s_{\bar qk}}+\frac{2z_q}{z_k s_{qk}}+\frac{z_b}{z_k s_{bk}}
\right)\right.\\
&\hspace{4cm}\left.+\,\frac{C_A}{2}\left(
\frac{2z_{\bar q}}{z_k s_{\bar q k}}-\frac{z_q}{z_ks_{qk}}-\frac{z_b}{z_ks_{bk}}-\frac{m_W^2}{s_{qk}s_{\bar q k}}-\frac{s_{b\bar q}}{s_{bk}s_{\bar q k}}+\frac{2s_{bq}}{s_{bk}s_{qk}}
\right)
\right]\nonumber\,.
\end{align}
Rather interestingly, this expression is {\it not} symmetric in the quark and anti-quark $q\leftrightarrow \bar q$.  The jet consists of two quarks ($b$ and $q$) and one anti-quark ($\bar q$), and so the flow of color around in the jet is not symmetric.  While one can consider the implications of this asymmetry, we will leave that for future work, and instead only consider the symmetrized matrix element, assuming complete ignorance as to which subjet is that of the quark or anti-quark.  The corresponding symmetrized soft gluon contribution is
\begin{align}
\frac{p_{b,\text{sym}}(\Pi_\text{NLO})}{p_{b,\text{sym}}(\Pi_\text{LO})}&\supset (4\pi)^2\left[
2C_F \left(
\frac{m_W^2}{s_{qk}s_{\bar qk}}+\frac{z_b}{z_k s_{bk}}
\right)\right.\\
&\hspace{3cm}\left.+\,\frac{C_A}{2}\left(
\frac{z_{\bar q}}{z_k s_{\bar q k}}+\frac{z_q}{z_ks_{qk}}+\frac{s_{b\bar q}}{s_{bk}s_{\bar q k}}+\frac{s_{bq}}{s_{bk}s_{qk}}-\frac{2z_b}{z_ks_{bk}}-\frac{2m_W^2}{s_{qk}s_{\bar q k}}
\right)
\right]\nonumber\,.
\end{align}

The difference between the background and signal distributions from soft emission at next-to-leading order that we consider here is then
\begin{align}
\frac{p_{b,\text{sym}}(\Pi_\text{NLO})}{p_{b,\text{sym}}(\Pi_\text{LO})}-\frac{p_s(\Pi_\text{NLO})}{p_s(\Pi_\text{LO})}&\supset  (4\pi)^2\frac{C_A}{2}
\left(
\frac{z_{\bar q}}{z_k s_{\bar q k}}+\frac{z_q}{z_ks_{qk}}+\frac{s_{b\bar q}}{s_{bk}s_{\bar q k}}+\frac{s_{bq}}{s_{bk}s_{qk}}-\frac{2z_b}{z_ks_{bk}}-\frac{2m_W^2}{s_{qk}s_{\bar q k}}
\right)\nonumber\\
&\hspace{6cm}+\text{non-singular}\,,
\end{align}
and note that the contribution proportional to the fundamental Casimir $C_F$ exactly cancels.  As mentioned earlier, one can take this expression and insert it into the general results for discrimination at next-to-leading order to calculate the ROC curve, but that will not be our approach here.  Instead, we will use this expression as inspiration for constructing a color-sensitive observable that can be practically used in a simulated or experimental analysis.

\subsection{An IRC Safe Color-Flow Sensitive Observable}

To this goal, let's first write down a more complete expression for the likelihood through next-to-leading order with these results.  We now have
\begin{align}\label{eq:nlolike}
{\cal L} = \frac{p_b(\Pi_\text{LO})}{p_s(\Pi_\text{LO})}\left[
1+\frac{\alpha_s}{2\pi}(4\pi)^2\frac{C_A}{2}
\left(
\frac{z_{\bar q}}{z_k s_{\bar q k}}+\frac{z_q}{z_ks_{qk}}+\frac{s_{b\bar q}}{s_{bk}s_{\bar q k}}+\frac{s_{bq}}{s_{bk}s_{qk}}-\frac{2z_b}{z_ks_{bk}}-\frac{2m_W^2}{s_{qk}s_{\bar q k}}
\right)+\cdots
\right]\,,
\end{align}
suppressing the normalization factor at next-to-leading order and non-singular terms.  Thus we see that the contribution to the likelihood at next-to-leading order can be positive or negative and correspondingly increases or decreases the likelihood from its leading-order value.  A negative contribution means that the soft gluon emission is more likely to have come from the color flow of signal, while a positive contribution means that it was more likely to have originated from the color flow of background.  The geometric dependence of the next-to-leading order contribution can be made more transparent by replacing the invariant masses with explicit pairwise angles, where $s_{ij} = z_iz_j E^2\theta_{ij}^2$, in the collinear limit, where $E$ is the jet energy.  With the appropriate replacements, the likelihood can be expressed as
\begin{align}\label{eq:angnlo}
{\cal L} = \frac{p_b(\Pi_\text{LO})}{p_s(\Pi_\text{LO})}\left[
1+\frac{\alpha_s}{2\pi}\frac{C_A}{2}\frac{(4\pi)^2}{z_k^2 E^2}
\left(
\frac{1}{\theta_{\bar q k}^2}+\frac{1}{\theta^2_{qk}}+\frac{\theta^2_{b\bar q}}{\theta^2_{bk}\theta^2_{\bar q k}}+\frac{\theta^2_{bq}}{\theta^2_{bk}\theta^2_{qk}}-\frac{2}{\theta^2_{bk}}-\frac{2\theta_{q\bar q}^2}{\theta^2_{qk}\theta^2_{\bar q k}}
\right)+\cdots
\right]\,.
\end{align}
Note that the phase space for the emission of a soft and collinear gluon is
\begin{align}
d\Pi_\text{soft-coll} = \frac{2E^2}{(4\pi)^3}\,d\theta^2\,z\, dz\, d\phi\,,
\end{align}
where $\theta^2$ is the polar angle, $\phi$ is the azimuthal angle, and $z$ is the energy fraction of the soft emission.  Note that factors of the jet energy $E^2$ and soft gluon energy fraction cancel between the expression of the likelihood and phase space.

From properties and features of this expression for the next-to-leading order likelihood, we would then like to construct an IRC safe observable that captures its features.  To be IRC safe, such an observable must vanish if the gluon has 0 energy, because we cannot in principle measure where such a gluon is emitted, and so we can write such an observable in the form
\begin{align}
{\cal O}_\text{NLO} = \sum_k z_k\, {\cal F}(\{\theta\})\,,
\end{align}
where ${\cal F}(\{\theta\})$ is a function of the various pairwise particle angles.  Here, we have also explicitly summed over all particles $k$ in the jet to anticipate its use in a practical analysis.  Na\"ively, we might think that the angular dependence of \Eq{eq:angnlo} could be copied directly as the function ${\cal F}(\{\theta\})$, but this has some undesirable properties.  In particular, just the angular dependence of \Eq{eq:angnlo} diverges in any collinear limit, $k\parallel b,q,\bar q$, but remains integrable because of the phase space measure.  

To construct a desirable observable, we will enumerate all vital features of the likelihood at next-to-leading order.  First, for collinear safety, the observable must vanish if the emission $k$ becomes collinear to any of the three hard, leading order particles, $k\parallel b,q,\bar q$.  Next, note that the net colors of the signal jet (the top quark) and the background jet (the bottom quark) are identical.  That is, if the soft gluon is emitted at sufficiently large angle so that it can only resolve the net color of the jet, signal and background appear identical.  Therefore, there is no discrimination power in the large angle limit, when $\theta_{bk},\theta_{qk},\theta_{\bar qk}\to \infty$.  In this limit, the observable should also vanish.  Next, the observable should be invariant to boosts along the jet axis, so that the observable itself imposes no absolute angular scales \cite{Larkoski:2013eya,Larkoski:2014gra}.  This can be easily accomplished by appropriate multiplication by factors of the energy of the jet.

The final property is sensitivity to the explicit form of the angular factor at next-to-leading order.  In particular, the fact that the sign of the angular factor varies on phase space tells you if the emission is more likely to be from signal or background.  Thus, we want to retain this geometric sign information.  To ensure this, we ensure that the function ${\cal F}(\{\theta\})$ has the exact same signs as the angular factor at next-to-leading order on phase space.  Note that the angular factor can be expressed over a common denominator as
\begin{align}
&\frac{1}{\theta_{\bar q k}^2}+\frac{1}{\theta^2_{qk}}+\frac{\theta^2_{b\bar q}}{\theta^2_{bk}\theta^2_{\bar q k}}+\frac{\theta^2_{bq}}{\theta^2_{bk}\theta^2_{qk}}-\frac{2}{\theta^2_{bk}}-\frac{2\theta_{q\bar q}^2}{\theta^2_{qk}\theta^2_{\bar q k}} \\
&\hspace{4cm}= \frac{\theta_{q k}^2\theta_{b\bar q}^2+\theta_{\bar q k}^2\theta_{bq}^2+\theta_{bk}^2\theta_{qk}^2+\theta_{bk}^2\theta_{\bar qk}^2-2\theta_{qk}^2\theta_{\bar q k}^2-2\theta_{q\bar q}^2\theta_{bk}^2}{\theta^2_{qk}\theta^2_{\bar q k}\theta^2_{bk}}\nonumber\,.
\end{align}
Thus, we require that
\begin{align}
{\cal F}(\{\theta\}) \propto \theta_{q k}^2\theta_{b\bar q}^2+\theta_{\bar q k}^2\theta_{bq}^2+\theta_{bk}^2\theta_{qk}^2+\theta_{bk}^2\theta_{\bar qk}^2-2\theta_{qk}^2\theta_{\bar q k}^2-2\theta_{q\bar q}^2\theta_{bk}^2\,.
\end{align}
Note also that this complicated angular factor vanishes in all collinear limits itself.

Then, to summarize, the properties we desire of the angular dependence of an IRC safe color flow-sensitive observable for top tagging are:
\begin{enumerate}
\item Vanishes in any collinear limit, $k\parallel b$, $k\parallel q$, or $k\parallel \bar q$.
\item Vanishes in the large angle limit, $\theta_{bk},\theta_{qk},\theta_{\bar qk}\to \infty$.
\item Invariant to boosts along the direction of the jet.
\item Proportional to $\theta_{q k}^2\theta_{b\bar q}^2+\theta_{\bar q k}^2\theta_{bq}^2+\theta_{bk}^2\theta_{qk}^2+\theta_{bk}^2\theta_{\bar qk}^2-2\theta_{qk}^2\theta_{\bar q k}^2-2\theta_{q\bar q}^2\theta_{bk}^2$.
\end{enumerate}
With these constraints, the form of the observable we consider here is then
\begin{align}\label{eq:nloobs}
{\cal O}_\text{NLO} \equiv \frac{m_t^2}{E^2}\sum_k z_k \,
\frac{\theta_{q k}^2\theta_{b\bar q}^2+\theta_{\bar q k}^2\theta_{bq}^2+\theta_{bk}^2\theta_{qk}^2+\theta_{bk}^2\theta_{\bar qk}^2-2\theta_{qk}^2\theta_{\bar q k}^2-2\theta_{q\bar q}^2\theta_{bk}^2}{\left(\theta_{bk}^2\theta_{qk}^2+\theta_{bk}^2\theta_{\bar qk}^2+\theta_{qk}^2\theta_{\bar qk}^2\right)^{3/2}}\,.
\end{align}
This observable satisfies all of the criteria, most of which are self-evident.  Boost invariance follows from noting that, in the collinear limit, angles scale like $\theta \to \gamma^{-1}\theta$ with boosts along the jet direction, where $\gamma$ is the boost factor, while the total jet energy scales like $E\to \gamma E$.  Note also that this form of the observable is not unique, and there may be benefit to considering different scaling with angles versus total jet energy.  However, we will just restrict our analysis to this form in this paper.

In practice, as we will study in the next section, we will consider the approximation to the likelihood observable through next-to-leading order constructed from the observables ${\cal O}_{\text{LO}}$ and ${\cal O}_{\text{NLO}}$ as
\begin{align}\label{eq:obsnlo}
{\cal O} \equiv {\cal O}_\text{LO}\left(
1+\alpha\, {\cal O}_{\text{NLO}}
\right)\,,
\end{align}
where $\alpha>0$ is a parameter on the scale of the coupling $\alpha_s$ that ensures that the observable ${\cal O}$ is always positive.  In particular, if $\alpha = 0$, then we will restrict the observable to its approximation to the leading-order likelihood ratio, and in general, $\alpha$ can be optimized in an analysis given pure signal and background distributions.

\begin{figure}[t]
\begin{center}
\includegraphics[width=0.45\textwidth]{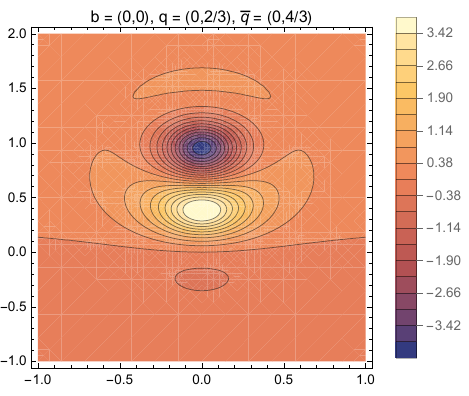}\ \ \ \includegraphics[width=0.45\textwidth]{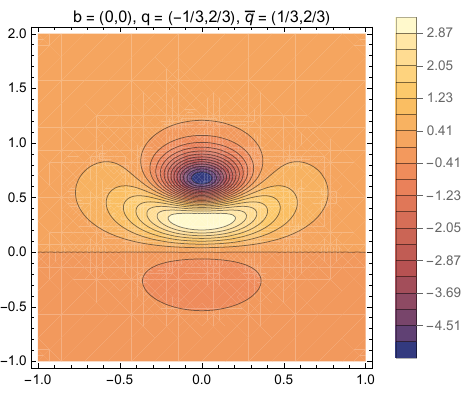}\\
\includegraphics[width=0.45\textwidth]{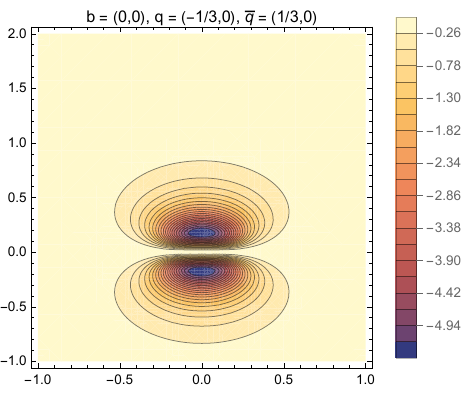}\ \ \ \includegraphics[width=0.45\textwidth]{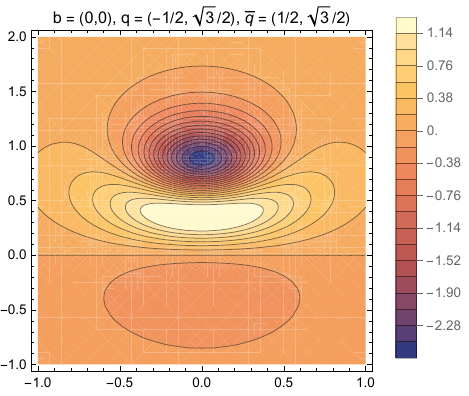}
\caption{\label{fig:geoconts}
Contour plots of the angular dependence of the approximate next-to-leading order likelihood observable, ${\cal O}_\text{NLO}$, with different choices of the location of the three hard particles at leading order.  In all plots, the bottom quark is located at the origin, $(0,0)$, while the quark and anti-quark are located at $q = (0,2/3)$, $\bar q = (0,4/3)$ (upper left), $q = (-1/3,2/3)$, $\bar q = (1/3,2/3)$ (upper right), $q = (-1/3,0)$, $\bar q = (1/3,0)$ (lower left), and $q = (-1/2,\sqrt{3}/2)$, $\bar q = (1/2,\sqrt{3}/2)$ (lower right).
}
\end{center}
\end{figure}

We present contour plots of the angular dependence of the observable ${\cal O}_\text{NLO}$ in \Fig{fig:geoconts}.  Here, we just plot the angular factor of \Eq{eq:obsnlo} on the collinear tangent plane of the celestial sphere or detector on which the jet is located.  In each plot, the location of the bottom quark is fixed to the origin, while the quark $q$ and anti-quark $\bar q$ are varied.  Negative values correspond to likely signal (top quark) origin, which is rather tightly isolated to the location between the $q$ and $\bar q$ (that is, from the $W$ boson decay).  By contrast, background is most likely to emit in the region between the bottom quark and the $q\bar q$ pair, which is a region that has no color connection in signal.  One interesting configuration is when the bottom quark lies between the $q$ and $\bar q$ (lower left in the figure), for which there is no region where the observable is positive, or no region likely corresponding to emission from background.  This configuration is rather special, but nevertheless illustrates how radiation is likely to be distributed.

\subsection{A Hard Limit to Discrimination Power}

A corollary to this soft gluon emission analysis at next-to-leading order is that there is no limit in which signal and background in this top tagging problem become perfectly distinct.  The likelihood observable through next-to-leading order in \Eq{eq:nlolike} only has support on a finite region of phase space, where both signal and background have a non-zero probability for emission.  This is to be contrasted with, for example, binary discrimination of boosted $H\to b\bar b$ decays from $g\to b\bar b$ fragmentation \cite{Buckley:2020kdp,Larkoski:2023xam}.  In the sufficiently highly-boosted limit, because the Higgs is a color-singlet, there is strictly 0 probability for gluon emission outside of the region of the $b\bar b$, while there is non-zero (and effectively unit) probability for the color-octet gluon to emit at wide angles.  Thus the (non-)existence of a single gluon emission outside the $b\bar b$ region tells you with perfect fidelity what particle initiated the jet, at least in the narrow-width approximation and assuming no contamination radiation.

For the problem of top tagging, because signal and background jets have the same net color, we expect that the efficacy of the likelihood observable converges rather quickly as a perturbative expansion in $\alpha_s$.  As we have observed, the kinematic distribution of the three hard subjets provides significant discrimination power at leading order, while sensitivity to the color flow beginning at next-to-leading order will not be able to provide much additional discrimination.  Additionally, the detailed flow of color and associated distribution of soft emissions may not be well-modeled in leading-order event generators and leading logarithmic parton showers, which typically work to leading order in the large number of colors $N_c$ limit anyway.  To ensure that this leading color sensitivity is correctly described would require matching to matrix elements at next-to-next-to-next-to-leading order (three emissions beyond leading order) with respect to the initial $pp\to b\bar b$ process, which is well beyond the accuracy of any extant program.  These are important to consider when drawing conclusions about optimal machine learning performance for this problem, that the simulated data that one works with at this extremely detailed level may simply be a poor approximation to nature.

That said, this analysis is limited to discrimination information that can be gained through the distribution of IRC safe energy flow and possibly quark flavor identification.  Top and bottom quarks have distinct electroweak quantum numbers and so sensitivity to non-IRC safe information may improve discrimination significantly.  Examples of such additional information are things like net electric charge of the jets, total hadronic multiplicity, or multiplicity of second generation hadrons \cite{Duarte-Campderros:2018ouv,Nakai:2020kuu,Albert:2022mpk,Bedeschi:2022rnj,Erdmann:2020ovh}.  The available discrimination power in these other quantities depends on the features and performance of one's detector, and so are not as universal in the same way as energy flow.  Nevertheless, their relevance for LHC physics is vital to study.  We leave such an analysis to future work.

\section{Top Tagging in Practice}\label{sec:toptagprac}

With these analytic predictions and expressions for useful observables, we now turn to establishing their utility in simulation.  We generated $pp\to t\bar t$ and $pp\to b\bar b$ events at the 14 TeV LHC in Pythia 8.240 \cite{Bierlich:2022pfr} with default settings, except forcing fully hadronic decays of the tops, and making all $b$-hadrons stable so that $b$-tagging could be implemented by particle identification.  Jets were found using FastJet 3.4.0 \cite{Cacciari:2011ma}, with the anti-$k_T$ algorithm  \cite{Cacciari:2008gp}, with a jet radius of $R = 0.5$.  We required that jets have a transverse momentum of greater than 1 TeV, $p_\perp > 1$ TeV, and whose center was located at less than an absolute value of 2.5 in pseudorapidity.  If multiple jets in the event pass this criteria, we select one at random.  Jets were then reclustered into three subjets with the exclusive $k_T$ algorithm \cite{Ellis:1993tq,Catani:1993hr}, with the Winner-Take-All recombination scheme \cite{Bertolini:2013iqa,Larkoski:2014uqa,gsalamwta}, that ensures that the subjet centers lie in the direction of hard particles in the subjets.  We demand that one of the subjets contains a $b$ hadron and record the total number of events that pass these criteria.  As additional kinematic constraints, we demand that the mass of the jet lies in the window $m_J \in[150,200]$ GeV, to be consistent with the top mass, and that the invariant mass of the two non-$b$ subjets is in the window $m_{q\bar q}\in[60,100]$ GeV, to be consistent with the $W$ mass.  We record the number of events that pass each of these mass cuts.  On the jets that remain after all these cuts, we record them for further analysis.

\begin{table}[t!]
\begin{center}
\begin{tabular}{c|cc|cc|}
 & \multicolumn{2}{|c|}{Signal} & \multicolumn{2}{|c|}{Background}\\
 \hline
$b$-tag, $p_\perp > 1$ TeV& 14791 events & 1 & 653297 events & 1 \\
$+\,m_J \in[150,200]$ GeV & 11480 events& 0.776  & 39908 events & 0.0611\\
  $+\,m_{q\bar q} \in[60,100]$ GeV & 9290 events & 0.628 & 8760 events & 0.0134\\
\end{tabular}
\end{center}
\caption{\label{tab:rates}
Table of number of events and corresponding fraction of total sample that passes cuts on signal (initial top quark jets) and background (initial bottom quark jets), as simulated in Pythia.  The first row lists the total number of jets in the ensemble that are $b$-tagged and have a transverse momentum greater than $1$ TeV.  The second row includes the constraint that the total jet mass is consistent with the top quark, while the third row includes the constraint that a pairwise invariant mass of subjets is consistent with the $W$ boson.
}
\end{table}

In \Tab{tab:rates}, we record the rates for these cuts as imposed on simulated signal and background events from Pythia.  As expected, additional total jet mass and pairwise subjet mass cuts have relatively small effect on top quark jets, decreasing the initial sample by less than 40\%.  By contrast, the mass cuts on background are catastrophic, reducing the initial sample by ultimately a factor of about 100.  This is a bit larger than the estimate of the analysis of \Sec{sec:nomratesest}, suggesting that the contribution of the $b\to bgg$ channel might be underestimated by a factor of 5 or so.  For this final state in particular, the jet radius and subjet clustering is important for IRC safety, and dependence on these parameters, which is not included in the estimate of \Sec{sec:nomratesest}, is likely the cause of the the difference.

On the jets that pass all of these cuts, we then measure the observables ${\cal O}_\text{LO}$ and ${\cal O}_\text{NLO}$, as defined earlier.  In this practical implementation at a hadron collider, we use longitudinally-invariant quantities, substituting transverse momentum with respect to the beam for energy, $E\to p_\perp$, and invariant cylindrical distance for angle:
\begin{align}
\theta_{ij}^2 \to (\eta_i-\eta_j)^2 + (\phi_i-\phi_j)^2\,,
\end{align}
where $\eta_i$ and $\phi_i$ are the pseudorapidity and azimuthal angle of particle $i$.  The direction of the three leading-order subjets are defined by their Winner-Take-All recombination scheme axes, from which pairwise angles are evaluated.  Using Winner-Take-All recombination ensures that the axes are recoil-free and whose direction is unaffected by soft, contamination radiation. 

\begin{figure}[t]
\begin{center}
\includegraphics[width=0.45\textwidth]{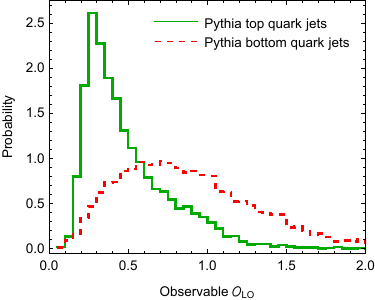} \ \ \ 
\includegraphics[width=0.45\textwidth]{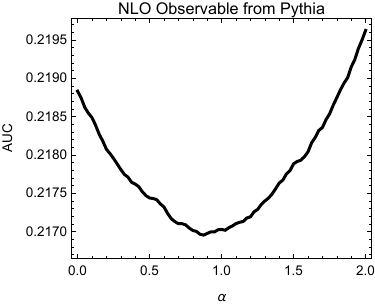}
\caption{\label{fig:pythiadists}
Left: Plot of the distribution of the observable ${\cal O}_\text{LO}$ on signal and background jets as generated with Pythia.  Right: Plot of the area under the ROC curve (AUC) as a function of the mixing parameter $\alpha$ in the observable ${\cal O} = {\cal O}_\text{LO}\left(
1+\alpha {\cal O}_\text{NLO}
\right)$ from signal and background jets from Pythia.  Smaller values correspond to better discrimination power.
}
\end{center}
\end{figure}

At left in \Fig{fig:pythiadists}, we plot the distribution of observable ${\cal O}_\text{LO}$ from \Eq{eq:loobs} on signal and background jets from Pythia.  These distributions are remarkably similar to our analytic predictions from \Sec{sec:lolike}, especially on signal jets, while the background distribution is, if anything, broader and more distinct from signal than that of the prediction.  This again is likely due to the other leading-order background process, $b\to bgg$, whose effect we hadn't included in our prediction.  Nevertheless, this demonstrates the importance and utility of a sound predictive framework in which results can be interpreted and understood.

Next, we then consider measuring the observable ${\cal O}$, where
\begin{align}
{\cal O} = {\cal O}_\text{LO}\left(
1+\alpha {\cal O}_\text{NLO}
\right)\,,
\end{align}
where ${\cal O}_\text{NLO}$ is defined in \Eq{eq:nloobs} and $\alpha>0$ is a parameter, as motivated by our analysis of the likelihood ratio through next-to-leading order.  As a function of $\alpha$, we can calculate the area under the ROC curve (AUC) for this observable and establish the value of $\alpha$ at which the AUC is minimized, corresponding to optimal discrimination power.  This scan is plotted at right in \Fig{fig:pythiadists} where we see that for $\alpha\sim 0.9$, the AUC is minimized.  The improvement of discrimination power from leading to next-to-leading order with this form of observable is rather small, only about a 1\% change in the AUC from the nominal $\alpha = 0$.  This could be partly due to a suboptimal expression for the observable ${\cal O}_\text{NLO}$, but is likely more due to the fact that the difference in the flow of color, and therefore the distribution of radiation at next-to-leading order, is minimally different in signal and background.  Both classes of jets are net color triplets, so wide-angle radiation provides no discrimination power, and only intricate geometric dependence on the precise angles from possible emitting dipoles within the jets provides some discrimination.  Another possibility is that this observable is actually significantly more discriminant, but the flow of color in background jets especially is simply not well-modeled in simulation at this high order.  This is a possibility, but to tease this out would require significant analytic work in evaluating and resumming the observable ${\cal O}_\text{NLO}$, or through generation of events and jets in a parton shower accurate to much higher logarithmic accuracy.  We therefore leave an understanding of this to future work.

\section{Conclusions}\label{sec:concs}

We presented a first analytical study of top quark tagging, through binary discrimination of boosted hadronically-decaying top quarks from bottom quark fragmentation, directly from the likelihood ratio as formed from fixed-order matrix elements.  This systematic construction motivated simple observables whose discrimination power was validated in simulated data and correspondingly provides a human interpretation to the output of machine learning studies for the same problem.  Our analysis suggests a discrimination floor, that beyond leading order in QCD perturbation theory there is little information that could be exploited for discrimination because, for example, the difference in the flow of color between signal and background is intricate and not necessarily well-modeled by standard simulation software.

Because the top quark decays to three particles at leading order, a description of the background process is highly sensitive to the approximations employed by the simulation.  This is especially important for machine learning studies, where the machine learns exclusively from the simulated data, but that may differ significantly from reality, or even just a systematic approximation to reality.  The effect and importance of numerous improvements in approximations, including matching to fixed-order at high multiplicity, including $1\to 3$ splitting functions, and incorporating subleading color contributions, may lead to significant modifications of extracted discrimination power and are, at the very least, required to understand systematic uncertainties from truncation of approximations in a machine learning study.

Even before such an analysis, we have provided a uniquely powerful observable for discrimination, \Eq{eq:loobs}, that can be readily implemented in any future analysis, on simulation or experimental data.  This observable has no parameters, has robust and theoretically-sound justification, and only requires identification of the three hard subjets in the purported top jet.  Phase space for top quark decay, after imposing top and $W$ mass constraints, is still 3 dimensional, and so constructing a useful form of such an observable exclusively from a machine learning analysis of simulated data, like, say, through the energy flow polynomials \cite{Komiske:2017aww}, would still require multiple terms whose pattern of coefficient structure may not be obvious.  However, very little human work was required to get to \Eq{eq:loobs}, and perhaps there are other problems that on the surface appear ripe for machine learning that are, after just a little thought, within the realm of human understanding.

\section*{Acknowledgements}

I thank Gregor Kasieczka for initiating my interest in this topic and Rikab Gambhir for detailed comments.

\bibliography{refs}

\end{document}